\documentclass[lineno]{article}

\usepackage[utf8]{inputenc}
\usepackage{graphicx}
\usepackage{amsfonts}
\usepackage[square,numbers]{natbib}
\usepackage{subcaption}
\usepackage{url,lineno}
\usepackage{tikz}
\usepackage{float}
\usepackage[normalem]{ulem}
\usepackage[labelfont=bf, font=scriptsize]{caption}
\usepackage{placeins}
\usepackage{authblk}
\usepackage[a4paper, total={6.3in, 10in}]{geometry}
\usepackage[colorlinks=true, citecolor=blue, filecolor=blue, linkcolor=black]{hyperref}
\usepackage{bbm}
\usepackage{color}
\usepackage{epstopdf}
\usepackage{stmaryrd}
\usepackage{comment}
\usepackage{multicol}
\usepackage{amsmath,amssymb} 
\usepackage{color}
\usepackage{mathtools}
\usepackage[T1]{fontenc}    
\usepackage{url}            
\usepackage{booktabs}       
\usepackage{amsfonts}       
\usepackage{nicefrac}       
\usepackage{microtype}      
\usepackage{lipsum}
\usepackage{xcolor,colortbl}
\usepackage{amsmath}
\usepackage[square,numbers]{natbib}
\usepackage{lastpage,fancyhdr}

\newcommand{\dint}{\mathrm{d}}

\newcommand{\be}{\begin{equation}}
\newcommand{\ee}{\end{equation}}
\newcommand{\bea}{\begin{eqnarray}}
\newcommand{\eea}{\end{eqnarray}}

\definecolor{lime}{HTML}{A6CE39}
\DeclareRobustCommand{\orcidicon}{%
	\begin{tikzpicture}
	\draw[lime, fill=lime] (0,0) 
	circle [radius=0.16] 
	node[white] {{\fontfamily{qag}\selectfont \tiny ID}};
	\draw[white, fill=white] (-0.0625,0.095) 
	circle [radius=0.007];
	\end{tikzpicture}
	\hspace{-2mm}
}

\foreach \x in {A, ..., Z}{%
	\expandafter\xdef\csname orcid\x\endcsname{\noexpand\href{https://orcid.org/\csname orcidauthor\x\endcsname}{\noexpand\orcidicon}}
}

\begin{document}  

\title{\textbf{Timeliness criticality in complex systems}}

\author[1,2]{Jos\'e Moran\orcidA{}}
\affil[1]{Mathematical Institute and Institute for New Economic Thinking at the Oxford Martin School, University of Oxford, Oxford, United Kingdom}
\affil[2]{Complexity Science Hub, Vienna,  Austria}
\author[3]{Matthijs Romeijnders\orcidB{}}
\affil[3]{Department of Information and Computing Sciences, Utrecht University, The Netherlands}
\author[4]{Pierre Le Doussal\orcidC{}}
\affil[4]{Laboratoire de Physique de l’\'Ecole Normale Sup\'erieure, ENS, Universit\'e PSL, CNRS, Sorbonne
Universit\'e, Universit\'e de Paris, France}
\author[5,6]{Frank P. Pijpers\orcidD{}}
\affil[5]{Statistics Netherlands, The Hague}
\affil[6]{Korteweg-de Vries Institute for Mathematics, University of Amsterdam, The Netherlands}
\author[7,8,9]{Utz Weitzel\orcidE{}}
\affil[7]{School of Business and Economics, Vrije Universiteit Amsterdam, Amsterdam, The Netherlands}
\affil[8] {Tinbergen Institute, Amsterdam, The Netherlands}
\affil[9] {Nijmegen School of Management, Radboud University, Nijmegen, The Netherlands}
\author[3,10]{Debabrata Panja\orcidF{}\footnote{Corresponding author: d.panja@uu.nl}}
\affil[10]{Centre for Complex Systems Studies, Utrecht University, The Netherlands}
\author[11,12]{Jean-Philippe Bouchaud\orcidG{}}
\affil[11]{Capital Fund Management, Paris, France}
\affil[12]{X-CFM Chair of EconophysiX, Ecole polytechnique, Palaiseau, France}


\maketitle 

\begin{abstract}
In complex systems, external parameters often determine the phase in which the system operates, i.e., its macroscopic behavior. For nearly a century, statistical physics has extensively studied systems' transitions across phases, (universal) critical exponents, and related dynamical properties. Here we consider the functionality of systems, notably operations in socio-technical ones, production in economic ones and, more generally, any schedule-based system, where timing is of crucial importance. We introduce a stylized model of delay propagation on temporal networks, where the magnitude of delay-mitigating buffer acts as a control parameter. The model exhibits {\it timeliness criticality}, a novel form of critical behavior. We characterize fluctuations near criticality, commonly referred to as ``avalanches'', and identify the corresponding critical exponents. The model exhibits timeliness criticality also when run on real-world temporal systems such as production networks. Additionally, we explore potential connections with the Mode-Coupling Theory of glasses, the depinning transition and the directed polymer problem.
\end{abstract}

\vspace{5mm}
The value of {\it timeliness\/} --- goods, services, or people being in the right place at the right time --- can hardly be overstated. For example, the value of a train ride or food greatly depreciates in the presence of delays. From the perspective of socio-technical systems (STSs), where technological and human elements interact to provide functional support to our societies, the user-centric examples above translate into the availability of crew, infrastructure and material in transport systems, or into the production, processing and distribution in food systems that must dynamically match demand. The concept of timeliness has been ubiquitously and integrally adopted as a quality standard across STSs, reaching beyond the provision of transport and food into all other input-output systems or processes that operate with interdependent events and breaks at particular time intervals (henceforth, {\it schedule-based systems}), such as healthcare systems, emergency response systems, computer and web service systems, or agricultural systems. In this definition, schedule-based systems do not need to be pre-determined, but can evolve dynamically, and they also do not need a central planner, but can develop decentrally.

There exist a variety of incentives for STS operators, often reinforced by competitive pressures, to increase cost- and time-efficiencies in order to achieve superior operational results \cite{Ourperspective_2023}. For example, train operators may have the goal to maximize the number of passengers to be transported by the network, i.e., to fit trains into ever tighter schedules. This can be achieved by reducing the {\it temporal buffer\/} that allows the system to absorb possible delays. Operators could, for example, shorten stopping times of trains in stations or reduce the amount of replacement crews on standby. In the extreme (hypothetical) case of zero temporal buffers in the whole system, the delay of an arriving train is directly copied to the departing train: clearly, any delay, however small, will reverberate through the entire system \cite{Panja2021, giannikas2022data}. The purpose of this paper is to demonstrate, in terms of a stylized model, that efficiency incentives give rise to a critical phenomenon we call {\it timeliness criticality}.

Our model can be extended from transport systems to other schedule-based types of STSs, including the propagation of supply fluctuations in production networks, where firms reduce costs by keeping the inventory of production inputs at a minimal level \cite{ledwoch2016systemic, brintrup2018supply}. Any delay in such a ``just in time'' supply chain can easily propagate down the production network~\cite{Colon2017}, with major disruptions, as exemplified by the Suez canal obstruction by a single container ship in 2021~\cite{AssociatedPress2021}. This is actually the same type of behavior one sees in transport systems, as inventories are conceptualized as buffers that allow firms to keep producing or selling in the absence of inputs for a given amount of time. The main difference with transport networks is that inventories are dynamic, and can be depleted after successive delay events if they are not properly replenished. Inventories can in fact help to substantially mitigate the economic effects of natural disasters~\cite{Colon2020}, but even in less extreme conditions, low levels of inventories or of supplier redundancy can cause strong output fluctuations of firm networks (see e.g. \cite{kosasih2022reinforcement, dessertaine} and references cited therein).

The aim of the present paper is to propose a minimal, stylized model that reveals the existence of a critical point in schedule-based systems as the size of the mitigating buffers is reduced. More precisely, above a certain critical buffer size, delays cannot propagate and large-scale system-wide delays are avoided. On the contrary, when temporal margins are not wide enough, delays accumulate without bounds. 

Our model is closely related to the so-called ``Bounded Kardar-Parisi-Zhang'' (B-KPZ) equation \cite{munoz1998nonlinear, hinrichsen2000non}, that describes the motion of a driven interface in the presence of a hard wall preventing the interface to visit the lower half space. Depending on the driving field, the interface is either localized around the hard wall, or undergoes an unbinding (or depinning) transition and escapes to infinity. Our model precisely describes such unbinding transition in the mean-field limit, where we find unexpected (and little understood at this stage) analogies with the Mode Coupling Theory of the glass transition \cite{gotze2009complex}.   

\section*{Timeliness criticality\label{sec2}}

Since our model is a representation of any form of planning schedule denoting temporal ordering of different events that (have to) depend on each other, 
it is best formulated in terms of {\it temporal networks\/} \cite{Holme2012,tempnetbook}. We describe the mathematical structure of our model below.
\begin{figure}[!h]
\begin{center}
\includegraphics[width=\linewidth]{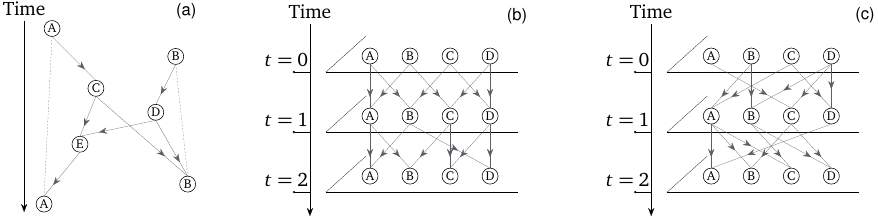}
    \caption{Schematics of our model in terms of temporal networks. (a) How the (time-ordered) dependencies --- shown by arrows --- among the system components (number of components $N=5$) in a real-world situation may play out [Eq.~\eqref{eq:def1}]. A mean-field (MF) (b) and a synthetic temporal network (STN) (c) representation respectively of our stylized model (number of components $N=4$) in discretized time steps [Eq.~\eqref{eq:em1}]. For both the MF and the STN cases, the number of arrows $K$ entering every system component is constant (e.g., $K=2$); for the latter case the number of arrows leaving every system component is also $K$.}
    \label{fig:fig1}
\end{center}
\end{figure}

\subsection*{Temporal networks\label{sec2a0}}

In general terms, the delay $\tau_i(t)$ suffered by a component $i \in \llbracket 1,N \rrbracket$ of the system at the $t$-th step of schedule, $t \in \mathbb{N}$, can be written as \cite{Panja2021}
\begin{equation}\label{eq:def1}
\tau_i(t) = \max_j\left[\sum_{t'=0}^{t-1} A_{ij}(t,t')\tau_j(t')- B_{ij}(t)\right]^+ + \varepsilon_i(t),
\end{equation}
where ${\mathbf A}(t,t')$ is the binary \textit{temporal adjacency matrix} of the network, indicating whether a delay of component $j$ at time $t'$ has an effect on the delay of component $i$ at time $t$, $B_{ij}(t)$ indicates the buffer that component $i$ has over component $j$ at time $t$, and $\varepsilon_i(t)$ is a random process encoding fluctuations affecting component $i$ at time $t$, with $[x]^+:=\max(0, x)$. Figure~\ref{fig:fig1}(a) provides a succinct representation of this structure for firm input-output production systems, for which every system component is a firm. The first term on the rhs can then be interpreted in terms of a firm $i$ whose production is affected at time $t$ because of delays on shipments from its suppliers $j$ that were sent out at an earlier time $t'$. Further, $B_{ij}$ is the {\it temporal buffer\/}: as explained in the introduction, it is a quantity of relevance in its own right for transport systems, and for firm input-output networks it is related to the amount of time firm $i$ can continue producing without being affected by a loss of inputs from firm $j$.

Equation~\eqref{eq:def1} is very general, and should in principle be accompanied by equations that specify the dynamics of the buffers $B_{ij}$ (for example as the inventories of firms get depleted) and of the adjacency matrix $\mathbf{A}$ (encoding firms that switch suppliers or transport vehicles and personnel that are rerouted, for example; that is to say that, in reality, a delay at time $t$ can impact the topology of the temporal network). Here we propose a much simpler framework. 

The first steps in simplifying Eq.~\eqref{eq:def1} are as follows: (a) assume that the dynamical evolution is ``Markovian'',
in the sense that the delay at layer $t$ only depends on events at $t-1$,
so that $A_{ij}(t,t')\equiv A_{ij}(t-1)\delta_{t',t-1}$, (b) we take $B_{ij}(t)=B$ to be constant and identical across all pairs of components.
Together these lead to the following stylized equation
\begin{equation}\label{eq:em1}
\tau_i(t) = \left[\max_{j} \left[A_{ij}(t-1)\tau_j(t-1)\right]-B\right]^+ +\varepsilon_{i}(t). 
\end{equation}

The adjacency matrix $A_{ij}(t)$ also remains to be defined. We work in the thermodynamic limit $N\to\infty$ and introduce two variants: (1) {\it mean-field\/} (MF), where a system component can be delayed by $K$ randomly chosen and statistically independent components that are re-drawn at every time-step $t$, meaning for all $t$, $\sum_{j}A_{ij}(t-1)=K$ [Fig.~\ref{fig:fig1}(b)], and (2) \textit{synthetic temporal network} (STN) that imposes further constraints, such as $\sum_{j}A_{ij}(t-1)=\sum_{j}A_{ji}(t)=K$ [Fig.~\ref{fig:fig1}(c)]. (Note that the indices $i,j$ are permuted in the two sums.) In the most general case, an event at time $t$ can be delayed by any event at a time $t'<t$. We restrict ourselves to the case where events can only affect events in the next time-step.

Finally, the statistics of the noise $\varepsilon_{i}$ need to be specified. In line with the Markovian assumption above, we take them to be iid (independent both in the component $i$ and time-step $t$) realizations of the same random variable. Only the MF case with exponentially-distributed noise allows us to find an analytical solution. 

For the sake of brevity, while we hold the example of the input-output network in mind, we use `node' to denote `system component' throughout the rest of the paper --- these are after all nodes in the concerned temporal networks.

\subsection*{Criticality \label{sec2a}}

An analytical solution of Eq.~\eqref{eq:em1} in the mean-field case can be provided if we take $\varepsilon$ to be an exponentially distributed random variable, with density $P(\varepsilon)=\mathbf{1}_{\varepsilon>0}\nu\exp(-\nu \varepsilon)$. Without any loss of generality we take $\nu=1$, choosing $\nu^{-1}$ as the unit for the buffer $B$ and the delays $\tau$.

We analyze $\psi_t(\tau)$, the distribution of the delays at time $t$ in the following manner. Having defined $\displaystyle{\Psi_t(\tau)}=\int_{\tau}^{\infty}\dint u~\psi_t(u)$, the complementary cumulative density function of the delays, and $Q_t(\tau)=K\psi_t(\tau)\left[1-\displaystyle{\Psi_t(\tau)}\right]^{K-1}$, the probability density of the maximum of $K$ randomly chosen delays, Eq.~\eqref{eq:em1} becomes (SI A)
\begin{equation}
\psi_t(\tau)\!=\!\!\int_0^\tau \!\!\!\dint \varepsilon~ P(\varepsilon)\,Q_{t-1}(\tau+B-\varepsilon) + P(\tau) \!\int_{0}^{B}\!\!\!\dint u ~Q_{t-1}(u),
\label{eq:em2}
\end{equation}
which is readily obtained by considering splitting Eq.~\eqref{eq:em1} into two cases: the first contribution corresponds to the maximal delay at the previous time-step being equal to $\tau+B-\varepsilon$ (i.e., larger than $B$ from which the buffer absorbs $B$ amount of delay), while the second one comes from the case when the maximum of the $K$ randomly chosen delays is smaller than $B$. With $P(\varepsilon)=\mathbf{1}_{\varepsilon>0}\exp(-\varepsilon)$, Eq.~\eqref{eq:em2} can be simplified to derive the following differential equation (SI A) 
\begin{eqnarray}
\Psi''_{t}(\tau)\!+\!\Psi'_{t}(\tau)\!\!\!&=&\!\!\! K[1\!-\!\Psi_{t-1}(\tau\!+\!B)]^{K-1}\,\Psi'_{t-1}(\tau\!+\!B).
  \label{eq:em3}
\end{eqnarray}

In the limit of large temporal depth $t\rightarrow\infty$,
we surmise (and check {\it a posteriori}) that the distribution of $\tau$ must have an exponential tail, leading to two different cases (SI B). In the first case, $\psi_t(\tau)$ and $\Psi_t(\tau)$ reach a stationary state asymptotically, with exponential tails $\propto\exp(-\alpha\tau)$ for some (inverse-time) parameter $\alpha$. This is a natural assumption, because the exponential distribution is in the Gumbel maximum universality class, and so the maximum of a large number of exponential random variables has a distribution with an exponential tail. Similarly, when $B\to\infty$ one simply has $\tau_i(t)=\varepsilon_i(t)$, and the delays are exponentially distributed with $\alpha =1$. This type of solution exists only when $B$ is larger than a non-trivial critical value $B^*_{\text c} \neq 0, \infty$, given by the condition (SI B)
\begin{eqnarray}
  B^*_{\text c}\exp(1-B^*_{\text c})=\frac1K,
  \label{eq:em4}
\end{eqnarray}
with an analytical solution $B^*_{\text c} = - W_{-1}\left(-1/(eK)\right)$, with $W_{-1}(x)$ the branch of the Lambert function with values in $[-1,-\infty[$ for $x \in [-1/e,0]$. In particular, this implies that $B_{\text c}^*\sim \log K$ asymptotically when $K \to \infty$.

In the second case, when $B<B^*_{\text c}$, the solution resembles that of a propagating front with ``velocity'' $v$, where asymptotically $\psi_t(\tau)\equiv\psi(\tau-vt)$ with $v=B^*_{\text c}-B$. Thus, $v=\mathbb{E}[\tau(t) -\tau(t-1)]$ corresponds to the mean delay accumulated between two successive iterations of the model.

These results imply that the {\it order parameter\/} for the development of delays in time, $v$, undergoes a {\it second-order phase transition\/} when the control parameter $B$ crosses $B^*_{\text c}$, with 
\begin{eqnarray}
v = \begin{cases}
  0 & \quad\mbox{if $B > B^*_{\text c}$}\\
  B^*_{\text c}-B & \quad \mbox{if $B < B^*_{\text c}$.}
  \end{cases}
\label{eq:em5}
\end{eqnarray}
\begin{figure}[!h]
  \centering
\includegraphics[width=\linewidth]{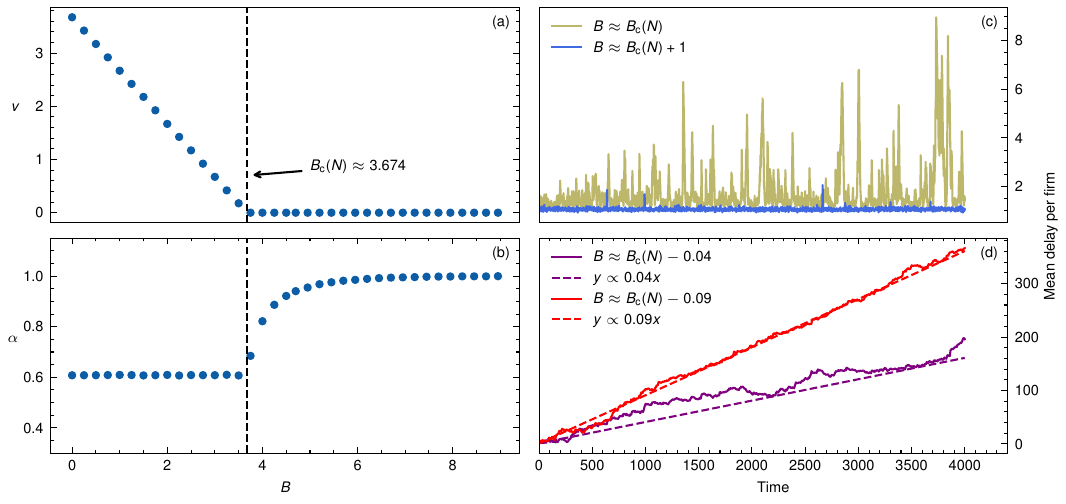}
\caption{Timeliness criticality. (a-b) Timeliness criticality in the stylized model obtained from direct simulations of Eq.~\eqref{eq:em1} with exponentially distributed noise for $N=10,000$ and $K=5$: both the order parameter $v$ and the exponent $\alpha$ for the exponentially decaying tail of the delay distribution function $\psi(\tau)$. The critical value of the temporal buffer $B_{\text c}(N)$ is found to be $\approx3.674$ (see text on how this value is obtained). (c-d) typical behavior of the mean delay per node $\sum_i \tau_i/N$, as a function of the temporal depth: when the buffer is smaller than the critical buffer $B_{\text c}(N)$ the mean delay per node keeps accumulating without bound (d); in contrast, large avalanches appear when the buffer is close to $B_{\text c}(N)$ from above, while a buffer significantly larger than  $B_{\text c}(N)$ results in small delays (c). The dashed lines in (d) show a fit through the data over a long period of time.}
  \label{fig:figure2}
\end{figure}

In particular, we obtain that $v(B=0)=B_{\text c}^*$, implying that the minimal buffer that must be applied to avoid any accumulation of delays corresponds precisely to the average delay accumulated in absence of buffers. Note also that $\alpha$ also has a different behavior above and below $B^*_{\text c}$, namely (SI B)
\begin{eqnarray}
\alpha = \begin{cases}
  1-\displaystyle{\frac{W_0(-B K e^{-B})}{B}} & \quad\mbox{if $B > B^*_{\text c}$}\\
  \alpha^*_{\text c} \equiv 1-\left(B^*_{\text c}\right)^{-1}& \quad \mbox{if $B < B^*_{\text c}$,}
  \end{cases}
\label{eq:alpha_c}
\end{eqnarray}
which has a square-root singularity $(B-B^*_{\text c})^{1/2}$ as $B\to B_{\text c}^*$ from above, suggesting that the transition is of the same kind as the K-core percolation transition, which itself is akin to the Mode-Coupling Theory of glasses \cite{sellitto2005facilitated, gotze2009complex}.  We coin the term {\it timeliness criticality\/} to describe this transition behavior. 

We used direct simulations of Eq.~\eqref{eq:em1} for various values of $N$ and $K$, in order to verify the above analytical results for the MF case. In Fig. \ref{fig:figure2}(a) we show the behavior of the order parameter $v$ as a function of the control parameter $B$ using exponentially distributed noise for $N=10,000$ and $K=5$ (we note that the STN case with exponential noise yields numerically indistinguishable results, although not shown here), from which we identify the critical buffer value by fitting a straight line of the form $v=B_{\text c}(N)-B$ for $B<B_{\text c}(N)$ following Eq. (\ref{eq:em5}). Correspondingly, the behavior of the parameter $\alpha$ in the exponential tail of $\psi(\tau)\equiv\psi_{t\rightarrow\infty}(\tau)$ is shown in Fig. \ref{fig:figure2}(b). In Figures \ref{fig:figure2}(c-d) we showcase the typical behavior of the mean delay per node: delays accumulate without bound when the buffer is smaller than the critical buffer $B_{\text c}(N)$ (d); in contrast, large avalanches of delays appear when the buffer is close to its critical value from above, while delays remain small when the buffer is substantially larger than its critical value (c).

From the comparison of Fig. \ref{fig:fig1}(a-b) and Table SI.1 in SI B, it is clear that the values of $B_{\text c}(N)$ and $\alpha_{\text c}(N)$ for $N=10,000$ and $K=5$ do not match the analytical solution $B^*_{\text c}\approx3.99431$ and $\alpha^*_{\text c}\approx0.749644$. We reason in SI B (Fig. SI.2 and the text above it) that this is caused by finite-$N$ effects. We also note that changing the distribution of the noise to, for example, a  half-Gaussian with $P(\varepsilon)=\mathbf{1}_{\varepsilon>0}\sqrt{2/\pi}\exp(-\varepsilon^2/2)$, yields qualitatively similar results. We expect the presence of a heavy (power-law) tail to change the nature of the transition but henceforth we limit ourselves only to exponentially distributed noise in this paper.

In the language of interfaces and of the B-KPZ equation, the delay accumulating transition reported above corresponds to a depinning transition \cite{munoz1998nonlinear, hinrichsen2000non}, which appear in many different contexts (domain walls in magnets, fracture fronts, yielding, etc.), see e.g. \cite{rosso2009avalanche, bouchaud1997scaling, lin2014scaling, ponson2017crack} and references therein.
\begin{figure}[!h]
\centering
\includegraphics[width=\linewidth]{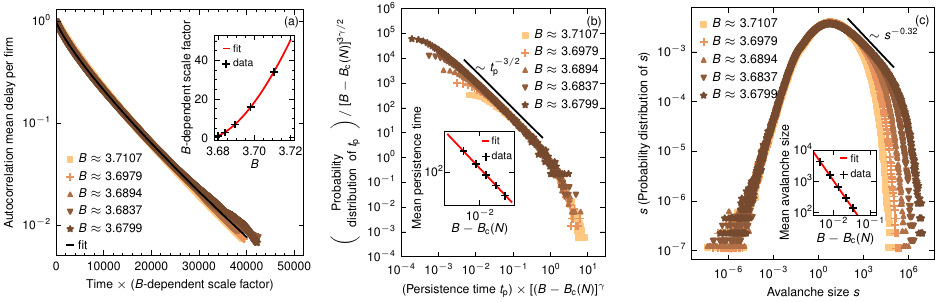}
\caption{Characterising timeliness criticality. (a) Autocorrelation function of mean delay per node for $N=10,000$ and $K=5$ for the MF case. The data have been collapsed to the reference curve  ($B\approx3.6799$) by stretching the $x$-axis by $B$-dependent scale factors. A stretched exponential of the form $\exp[-(t/t_{\text{ref}})^\beta]$ in temporal depth $t$, with best-fitted values $t_{\text{ref}}\approx5835.6$ and $\beta \approx0.823$ provides an excellent fit to the collapsed data. Inset: these scale factors are fitted by the function \eqref{eq:em7}, showcasing that the correlation length in the temporal domain diverges as a power-law when $B$ approached $B^*_{\text c}$ from above; see text for details. Probability distribution of avalanche persistence times (b), and of avalanche sizes (c) for the MF case, with $N=10,000$ and $K=5$ (see Methods for the definitions of these quantities). Panel (b): data collapse for the probability distribution of avalanche persistence times by scaling the $x$- and the $y$-axes by $[B-B_{\text c}(N)]^{-\gamma}$ and $[B-B_{\text c}(N)]^{-3\gamma/2}$ respectively, with the $B_{\text c}(N)\approx3.6755$ and $\beta\approx1.6936$, determined from panel (a). Inset (b): mean persistence times for the five $B$-values shown in the main plot;  the red line corresponds to $\approx2.355[B-B_{\text c}(N)]^{-\gamma/2}$, implying that the mean avalanche persistence time diverges $\sim[B-B_{\text c}(N)]^{-\gamma/2}$ (as it must given the data collapse of the main plot) when $B\rightarrow B_{\text c}(N)$ from above. Panel (c): the probability distribution of avalanche sizes is independent of $B$ at the lower end of the avalanche size, but at the higher end it behaves with an (apparent) power-law with exponent $\approx-0.32$, with the power law lasting deeper into the avalanche size as $B\rightarrow B_{\text c}(N)$ from above. Note that each tick on the x-axis corresponds to 3 decades. Inset (c): mean avalanche sizes for the five $B$-values shown in the main plot; the red line $\approx0.443[B-B_{\text c}(N)]^{-\gamma}$ is a numerically obtained fit to the data, implying that the mean avalanche size diverges $\sim[B-B_{\text c}(N)]^{-\gamma}$ as $B\rightarrow B_{\text c}(N)$ from above. See also main text.}
\label{avalanche}
\end{figure}
\section*{Characterizing timeliness criticality\label{sec3}}

\subsection*{Temporal correlations\label{sec3a}}

In critical phenomena one expects (power-law) diverging correlation lengths as the control parameter gets closer to the critical value in the disordered phase. For timeliness criticality, we expect diverging correlation lengths \textit{in time} as $B\rightarrow B^*_{\text c}$ from above. There is however a subtlety: correlations in the delays of \textit{individual nodes} decorrelate within times of $O(1)$. For the MF case, to which we confine ourselves in this section, this should however not come as a surprise since the temporal adjacency matrix ${\mathbf A}$ is composed randomly at every time step. 

That said, from Figs. \ref{fig:figure2}(c-d) we expect the signs of diverging correlation lengths to be picked up by the mean delay per node, defined as $\sum_i \tau_i/N$. This is explored in Fig. \ref{avalanche}(a) for $N=10,000$ and $K=5$, where we collapse the temporal  autocorrelation function of the mean delay per node by stretching the $x$-axis by $B$-dependent scale factors for five different $B$-values. These scale factors are then plotted in the inset of Fig. \ref{avalanche}(a) (symbols) and fitted with a curve 
\begin{eqnarray}
B\mbox{-dependent scale factor}=\frac{\left[B-B_{\text c}(N)\right]^\gamma}{\left[B_{\text{ref}}-B_{\text c}(N)\right]^\gamma},
\label{eq:em7}
\end{eqnarray}
which we have (arbitrarily) normalized  
using $B_{\text{ref}}=3.6799$. From such a fit
we obtain $B_{\text c}(N)\approx3.6755$ and $\gamma \approx1.6936$. In other words, Fig. \ref{avalanche}(a) showcases that the correlation time indeed diverges as a power-law when $B$ approaches the critical value from above. 

The analysis follows a standard method used for identifying critical exponents with a high degree of numerical precision~\cite{Newman_Barkema_book}. We note however that the collapse we obtain is very sensitive to the fitting parameters  $B_{\text c}(N)$ and $\gamma$ (in particular to the former). We also note that the shape of the collapsed curves in Fig. \ref{avalanche}(a) suggests that the correlation functions do not decay as a pure exponential: numerically, we find that a stretched exponential, with a stretching exponent $\beta \approx0.823$, produces an excellent fit to the collapsed curve.

As we have seen above, exponent $\gamma$, governing the divergence of the relaxation time as $B_{\text c}^*$ is approached, is rather precisely determined by our numerical data. However, perhaps surprisingly in view of the usual universality property of critical exponents, we have found that $\gamma$ depends on, e.g., the connectivity of the network $K$. For example, the data collapse of the autocorrelation of mean delay per node for $N=10,000$ and $K=7$ is also excellent and yields $B_{\text c}(N)\approx4.0441$ and $\gamma\approx1.745$, distinctly different from the value $\gamma\approx 1.6936$ obtained for $K=5$. Other exponents, such as stretching exponent $\beta$ or the exponent governing the avalanche persistence time reported in the next section, {\it are} universal. Such a coexistence of universal and non universal exponents is, as already suggested above, reminiscent of a similar situation for the Mode-Coupling Theory of glasses \cite{gotze2009complex}. There too, the exponent $\gamma$ describing the divergence of the relaxation time as the glass transition is approached is non universal, whereas the Debye-Waller factor has a universal square-root singularity in the glass phase. 

\subsection*{Delay avalanches}

From Fig. \ref{fig:figure2}(c) it is imperative that we analyze delay ``avalanches'', in close analogy with avalanches appearing at depinning, fracture or yielding transitions \cite{rosso2009avalanche, ponson2017crack, lin2014scaling}. Since delays can never be negative, delays that are above a certain threshold are of particular interest. In fact, the buffer $B$ is such a natural threshold, since delays below $B$ at a given time step gets absorbed by the buffer. Based on Figs. \ref{fig:figure2}(c-d) we therefore again consider mean delay per node, and define as avalanche those cases when the mean delay per node is $>B$. This brings two aspects of avalanches in focus: (i) {\it avalanche persistence times\/}, and (ii) {\it avalanche size\/}. How we measure these quantities are explained and illustrated in the Methods section.

The probability distributions of persistence times and avalanches sizes for $N=10,000$ and $K=5$ for the MF case are shown in Figs. \ref{avalanche}(a) and (b) respectively. Using $B_{\text c}(N)\approx3.6755$ and $\gamma\approx1.6936$ as found in Fig. \ref{avalanche}(b) we obtain an excellent collapse for the probability distributions of the persistence times: Fig. \ref{avalanche}(a) shows that long persistence times become increasingly likely as a power-law with an exponent $\approx-3/2$, corresponding to the return time probability density of an unbiased random walk. Further, the data collapse obtained by scaling the $x$-axis by $[B-B_{\text c}(N)]^{-\gamma}$ means that the {\it mean avalanche time} diverges more slowly, as $[B-B_{\text c}(N)]^{-\gamma/2}$, shown in the inset.

Similarly, large avalanches also become increasingly likely as a power-law with an (apparent) exponent $\approx-0.32$, with the power-law behavior holding on for progressively larger avalanche sizes as $B\rightarrow B_{\text c}$ from above, although the probability distribution of avalanche sizes remains independent of the value of $B$ [Fig. \ref{avalanche}(c)]. Finally, using a numerical fit to the data the mean avalanche size is seen to increase $\sim[B-B_{\text c}(N)]^{-\gamma}$ [inset of Fig. \ref{avalanche}(c)]. We note that these results are in agreement with the qualitative picture of Fig. \ref{fig:figure2}(c-d). We also note that the fluctuations in mean delay per node have been analysed in SI B (specifically, Fig. SI.3 and the text around it).

\section*{Timeliness criticality on real-world temporal networks \label{sec4}}

In contrast to the stylized models we have analyzed so far, in temporal networks describing real-world processes [e.g., Fig. \ref{fig:fig1}(a)], delays do not develop at discrete time steps. Real firms are not located on a ``lattice'' either, as the one shown in Figs. \ref{fig:fig1}(b-c), nor are the buffers of the same magnitude ($=B$) everywhere. Therefore, in order to bridge the gap between stylized models and the real world, we investigate the development of delays on two real-world temporal networks and demonstrate the proof-of-principle applicability of the timeliness criticality concept to the real world.
\begin{figure}[!h]
\centering
\includegraphics[width=\linewidth]{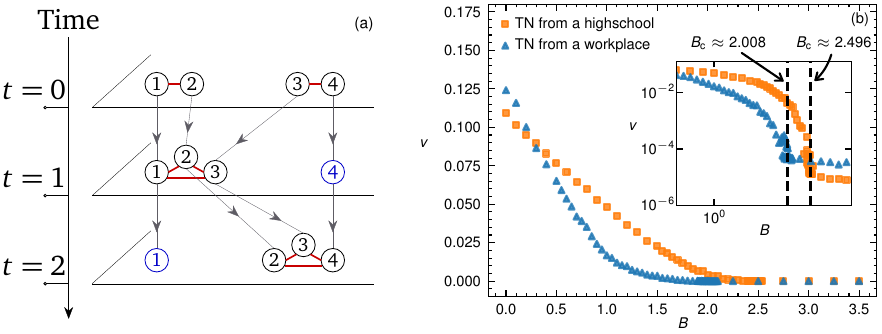}
\caption{Timeliness criticality on real-world temporal networks. (a) An example event representation for a real-world temporal network, consisting of four agents, numbered 1 through 4. The contact links among the agents at any time step is denoted by a thick red line, and each connected component --- such as 1-2 and 3-4 at time step 0 constitutes an event. An event can consist of a single agent as well, denoted by a blue agent node and blue outgoing connections. Single-agent events do not propagate delay according to Eq. \ref{eq:em1}, because they miss the noise term $\epsilon_i(t)$. Once the events are constructed at all time steps, agents can be seen to move from one event to another across time steps. (b) The $v$ vs $B$ graph for the two real-world temporal networks (TNs), the inset showing the same data as in the main graph, but with double logarithmic axes. At the $B$-values noted in the inset by dashed lines, the slopes of $\log v$ sharply approach infinity; these we identify as the critical buffer $B_{\text c}$. One can fit the singularity at $B_{\text c}$ as a power-law $(B_{\text c}-B)^\zeta$ with $\zeta \approx 5/2$.}
    \label{fig:RW_v_graph}
\end{figure}
Many real-world operational data are often too coarse or not easily accessible. Firm-to-firm production network data, for example, are often not available at a very detailed level, do not usually contain data on inventories, and/or are proprietary data. We therefore choose temporally-resolved contact data among pupils from a high school \cite{highschooldata} and employees at a workplace \cite{workplacedata}, both of which datasets are  publicly available from \url{http://www.sociopatterns.org} (data description in SI C). The contacts are traced at discrete time steps (at 20 seconds interval), and the temporal network is highly disordered, i.e., there are large variations in the temporal adjacency matrix ${\mathbf A}(t)$, which are directly taken from the empirical data. 

We ``event mapped'' \cite{Panja2022,Panja2022a} these temporal networks to suit our purpose: an example is shown in Fig. \ref{fig:RW_v_graph}(a), wherein contacts among individual agents (pupils, employees; each agent with a unique alphanumeric tag) at discrete time steps are shown by thick red lines. Every connected component of the network is termed ``an event''; once the events at every time step is constructed, in the temporal networks the agents can be seen as moving from one event to another across time steps. Note that the number of agents entering a specific event must also be equal to the number of agents leaving that event, albeit, unlike the case of STNs, the number of agents per event is not a constant (neither is the number of events taking place per time step). 

We simulated delay development [Eq.~\eqref{eq:em1}] on these temporal networks for uniform buffer $B$. Delays develop in a manner analogous to Fig. \ref{fig:fig1}(b): delays are carried by agents, and the delay of an event $i$ at time $t$ is the maximum of the delays of the agents entering that event plus an exponentially distributed random noise term $\varepsilon_i$ with exponent unity. Note however the presence of single-agent ``events'' [colored blue in Fig. \ref{fig:RW_v_graph}(b)]: for such events no noise term is added, as they are considered to be temporarily outside the system (e.g., a transport crew having a lunch break).

The corresponding $v$ vs $B$ plots, analog of Fig. \ref{fig:figure2}(a) are shown in Fig. \ref{fig:RW_v_graph}(b) (main graph), and with a logarithmic $y$-axis (inset). Evidently, there are critical transitions for both datasets, as the $v$ vs $B$ curves exhibit a power-law singularity (with a power larger than $1$) in $(B_{\text c}-B)$, signified by $d(\log v)/d(\log B)$ approaching infinity from the left, allowing us to cleanly identify the corresponding critical values $B_{\text c}$ [inset, Fig. \ref{fig:RW_v_graph}(b)]. In this context, we note that disorder increases the order of transition has been observed elsewhere \cite{Roux1987, lin2014scaling}. 

In contrast to Fig. \ref{fig:figure2}(a), the transitions are however not only smooth in $v$ vs. $B$ plots [e.g., $v$ does not have a discontinuous first derivative in $B$ as in Fig. \ref{fig:figure2}(a)], but we also find them taking place at lower values of $B_{\text c}$ than expected for STNs with similar numbers of agents. In SI D we argue that this is likely caused by the ``sparsity'' rather than by the ``heterogeneity'' in the temporal network topology (the terms are defined in SI D). 

\section*{Conclusion and  Discussions\label{sec5}}

The value of goods and services depends on the timeliness of their provisions; for example, the value of a ride service is not only to go from A to B but to also do so on time. Similarly, the economic value of food or a glass of water is also determined by whether they are available on time. System operators that provide these goods and services, on the one hand, have therefore integrally adopted timeliness as a quality standard, and on the other hand, are continuously striving to improve cost- and time-efficiencies in order to achieve superior operational results. We have captured the competition between achieving timeliness and striving for efficiency -- and more generally between efficiency and resilience \cite{Ourperspective_2023} -- in terms of a stylized model on temporal networks with a delay-mitigating temporal buffer $B$: more buffer translates to higher inefficiency, but more adherence to timeliness and stability. With $B$ as the control parameter, the model exhibits {\it timeliness criticality}, a novel form of phase transition {\it in time\/} that occurs at a critical value of $B$. Above this critical value, delay avalanches of all sizes appear spontaneously. For the stylized model we have characterized these avalanches and have identified the corresponding critical exponents. 

We have also shown that our model run on real-world temporal networks, which are not regular lattices, also exhibits timeliness criticality. Nevertheless, several challenges still exist for a meaningful translation of temporal buffers in our model to the real world, in particular for the case of production networks. First, temporal buffers are an aggregate of diverse measures in reality. For example, a production planner in a firm can increase temporal buffers by producing faster or in overtime, by increasing inventories, securing more suppliers for the same input, improving (electronic) procurement systems, changing product designs to reduce exposure to critical inputs, and so on \cite{brintrup2018supply}. Each of these measures comes with its own possibilities and limitations in terms of efficiency and buffering dynamics. Second, many firms have extensive production planning and operations research departments that tackle the trade-off between cost efficiency and delivery security on time \cite{kosasih2022reinforcement}. However, this research and planning is mostly limited to units within the firm, and to other firms with direct contractual links (suppliers and customers). The supplier of the supplier is already less foreseeable, not to speak of a supplier three or four links away on another continent \cite{wichmann2018towards}. Lastly, an important element that our simplified model does not capture is the heterogeneity in time delays and inventory levels, a point that was shown to be important in the model described in~\cite{Colon2017}. The complex interactions among firms, simultaneously trying to anticipate the actions of other firms, can lead to emergent system dynamics, as demonstrated by the `bullwhip effect', illustrated through the well-known `beer game`~\cite{Sterman1989}. Hence, both the heterogeneous composition of temporal buffers and their myopic adjustment can have very different implications for emergent behavior and timeliness criticality.

Of particular interest for the applicability of our model to real STSs is how relatively small and/or local events can lead to avalanches and system-wide disruptions, such as the cancellation of all train rides to reboot scheduling \cite{Panja2021a}, or a worldwide supply chain blockage due to natural disasters \cite{Shughrue2020}, or even a full-blown economic crisis. Following the early work of Bak et al. \cite{bak1993aggregate}, one might argue, as in \cite{moran}, that the tendency of efficiency-driven operators to self-organize into criticality explains why economies are more volatile than expected based on economic equilibrium models with rational expectations. This well-known phenomenon is referred to as the ``small shocks, large business cycle puzzle'' \cite{Bernanke1996,firefighting} -- the explanation of which, despite numerous attempts, remains elusive (see e.g. \cite{gabaix, dessertaine} and \cite{carvalho} for a recent review). The ``timeliness criticality'' developing on firm-to-firm networks may be a potent cause for an imbalance in demand and supply, leading to large scale volatility in economic output, as also argued in \cite{Colon2017}. This is in fact very visible in agent-based economic models~\cite{Colon2020,dessertaine,Pichler2021}, and aligned with empirical evidence of firms with low inventories being less resilient in crises~\cite{LafrogneJoussier2022}. Being able to apply such modelling to the real-world in an effective way requires, as highlighted above, much larger amounts of data than what is currently available, and therefore a large-scale effort of data collection~\cite{sc_alliance}. 

Going beyond STSs, the concept of timeliness criticality may be a promising approach for other schedule-based systems with incentives for improving efficiency like, e.g. distributed computing systems or task schedulers. Schedule-based systems do not need to be pre-determined and can evolve dynamically in a decentralized way and without a central planner. The framework introduced here applies in such cases too, and so it can potentially be applied to any input-output system with interdependent events at particular time intervals, for example, in biology between species, bodies, or cells.

\vspace{3mm}
\noindent {\bf Author contributions.} J-PB and DP conceptualized timeliness criticality. J-PB, PLD, JM, DP, FPP and MR analyzed timeliness criticality. All authors contributed to the manuscript.

\vspace{3mm}
\noindent{\bf Competing interests.} The authors have no competing interests to declare.

\section*{Methods}

\noindent {\bf Delay avalanches.} An example of delay development in the system has been shown in Fig. \ref{fig:figure2}(c) as the mean delay per node. We define the time that the mean delay per node remains consistently above $B$ as an avalanche. This is illustrated in Extended Data Fig. 1 for the mean-field case for $N=10,000$, $K=7$ [$B_{\text c}(N)\approx4.0435$ for these values] and $B\approx4.0503$, where we plot the mean delay per node as a function of time. We see that over the time interval $[t_{\text{start}},t_{\text{end}})$ with $t_{\text{start}}=4,760,908$ and $t_{\text{end}}=4,785,490$ the mean delay per node consistently stays above $B$, constituting an example delay avalanche, and the mean delay plot between these times is the ``avalanche curve''. The persistence time for the avalanche is then $t_{\text p}=t_{\text{end}}-t_{\text{start}}=24,582$, while the avalanche size is the area between the avalanche curve and the horizontal line denoting the value of $B$ over the interval $[t_{\text{start}},t_{\text{end}})$.
\setcounter{figure}{0}
\renewcommand{\figurename}{Extended data Figure}
\begin{figure}[!h]
\centering
\includegraphics[width=0.5\linewidth]{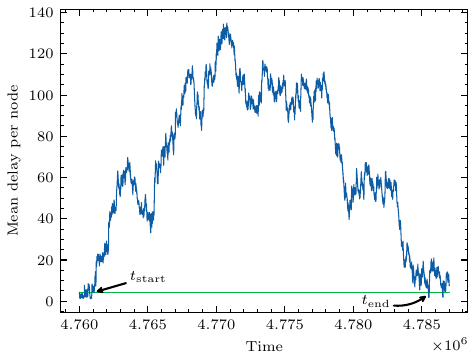}
\caption{An example of how an avalanche is defined. The mean delay per node for the mean-field case is plotted (blue), along with an orange line corresponding to mean delay equals $B$ (see text for the parameter values used for this plot). For the avalanche shown, we define $t_{\text{start}}=4,760,908$ as the time when the mean delay per node exceeds $B$, and $t_{\text{end}}=4,785,490$ as the time when for the first time after $t_{\text{start}}$ the mean delay per node falls below $B$ again. Then the mean delay per node in the interval $[t_{\text{start}},t_{\text{end}})$ constitutes a delay avalanche, from which the persistence time and the size of the avalanche are obtained respectively as $t_{\text p}=t_{\text{end}}-t_{\text{start}}=24,582$, and as the area between the blue curve and the orange line over the interval $[t_{\text{start}},t_{\text{end}})$.}
\label{fig:single_av}
\end{figure}

\vspace{5mm}
\noindent {\bf Data availability.} The real-world temporal networks data used in the paper are publicly available \cite{highschooldata,workplacedata}. The source data for all figures in the paper (including those
in the SI) are linked from the ``Data and codes for figures'' section in the SI to the respective folders in
\url{https://github.com/jose-moran/timeliness_criticality/tree/main/figures/}.

\vspace{5mm}
\noindent {\bf Code availability.} The codes used for generating results in the paper are available at \url{https://github.com/jose-moran/timeliness_criticality}. The codes to generate the figures from the source data are linked from the ``Data and codes for figures'' section in the SI to the respective folders in \url{https://github.com/jose-moran/timeliness_criticality/tree/main/figures/}.

\end{document}



\renewcommand{\theequation}{SI.\arabic{equation}}

\renewcommand{\thefigure}{SI.\arabic{figure}}

\renewcommand{\thetable}{SI.\arabic{table}}
\vfill
\tableofcontents
\vfill
\newpage
\section*{SI A: Evolution of delays for exponentially distributed noise}\label{sec:SIA}

\addcontentsline{toc}{section}{\protect\numberline{}SI A: Evolution of delays for exponentially distributed noise}

\noindent We assume that the delay $\varepsilon_i(t)$ is an i.i.d. random variable, taken from the distribution $P(\varepsilon)=\mathbf{1}_{\varepsilon>0}\exp(-\varepsilon)$. Let us denote the probability distribution function (pdf) of the delays at the time step $t$ by $\psi_t(\tau)$, with $\psi_0(\tau)=P(\tau)$. 

\vspace{3mm} 
\noindent The way to construct $\psi_{t+1}$ from $\psi_t$ is by taking $K$ samples from the pdf $\psi_t$. As explained in Eq. (3), at time step $t+1$ production at a firm can have an output delay $\tau$ in two distinct ways: (i) when the max of the $K$ chosen samples from $\psi_t$ have a delay equal to $\tau+B-\varepsilon$, and (ii) when the max of the $K$ chosen samples from $\psi_t$ have a delay $<B$. In the first case, $B$ amount of delay gets absorbed from the maximum and $\varepsilon$ gets added to the delay of from the noise term. Then we need to integrate the noise term $\varepsilon$ from $0$ to $\tau$. In the second case, $\psi_{t+1}(\tau)\sim\exp(- \tau)$, inherited from the distribution of the noise itself. In other words,
\begin{eqnarray}
  \psi_{t+1}(\tau)=\int_0^\tau
  \dint\varepsilon~\exp(-\varepsilon)\underbrace{[\mbox{pdf of max of the $K$
  chosen samples has value
  $(\tau+B-\varepsilon)$]}}_{Q(\tau+B-\varepsilon)}\nonumber\\&&
       \hspace{-12.4cm}+\exp(- \tau)\int_0^B \dint u~\underbrace{[\mbox{pdf of max of the $K$
  chosen samples has value
  $u$]}}_{Q(u)}
  \label{eq:e2}
\end{eqnarray}

\vspace{3mm} \noindent We choose $K$ samples from the probability distribution to satisfy that the maximum of them equals $u$. So we choose one of them with a delay $u$, and the rest $(K-1)$ samples have delays $\leq u$. Moreover, the one with a delay $u$ can be chosen in $K$ distinct ways, meaning that
\begin{eqnarray}
  Q(u)=K\,\psi_t(u)\,[1-\Psi_t(u)]^{K-1},
  \label{eq:e3}
\end{eqnarray}
where $\Psi_t(u)$ given by
\begin{eqnarray}
 \Psi_t(u)=\int_u^\infty \dint u'~\psi_t(u'),
  \label{eq:e4}
\end{eqnarray}
leading to
\begin{eqnarray}
 \psi_t(u)=-\frac{\dint}{\dint u}\Psi_t(u).
  \label{eq:e5}
\end{eqnarray}
This means
\begin{eqnarray}
  Q(u) = \frac{\dint }{\dint u}[1-\Psi_t(u)]^K\quad\mbox{and}\quad Q(\tau+B-\varepsilon)=-\frac{\dint }{\dint \varepsilon} [1-\Psi_t(\tau+B-\varepsilon)]^K.
  \label{eq:e6}
\end{eqnarray}

\vspace{3mm}
\noindent Using Eq.~\eqref{eq:e6} in Eq.~\eqref{eq:e2} we get
\begin{eqnarray}
\psi_{t+1}(\tau)=\exp(- \tau)\!\!\!\underbrace{\int_0^B \dint u\,\frac{\dint}{\dint u}[1-\Psi_t(u)]^K}_{=\,[1-\Psi_t(B)]^K,\,\text{since $\Psi_t(0)=1$}}-\int_0^d \dint \varepsilon~\exp(-\varepsilon)\,\frac{\dint}{\dint \varepsilon} [1-\Psi_t(\tau+B-\varepsilon)]^K.
  \label{eq:e7}
\end{eqnarray}
We do an integration by parts of the second term to write 
\begin{eqnarray}
\psi_{t+1}(\tau)\!\!&=&\!\! [1-\Psi_t(B)]^K\exp(- \tau)-\exp(-\varepsilon)\,[1-\Psi_t(\tau+B-\varepsilon)]^K\bigg|_{\varepsilon=0}^{\varepsilon=\tau}\nonumber\\&&\hspace{0cm}-\,\int_0^\tau
\dint \varepsilon~\exp(-\varepsilon)\,[1-\Psi_t(\tau+B-\varepsilon)]^K\!\!\nonumber\\
&=&\!\! [1-\Psi_t(\tau+B)]^K-\!\!\int_0^\tau
\!\!\dint u~\exp[-(\tau-u)]\,[1-\Psi_t(u+B)]^K.
\label{eq:e8}
\end{eqnarray}

\vspace{3mm}
\noindent A first order derivative of Eq.~\eqref{eq:e8} wrt $\tau$ yields
\begin{eqnarray}
\psi'_{t+1}(\tau)\!\!&=&\!\!-
K[1-\Psi_t(\tau+B)]^{K-1}\,\Psi'_t(\tau+B)
\nonumber\\&&-\underbrace{\left[[1-\Psi_t(\tau+B)]^K-\!\!\int_0^\tau\!\!\dint u~\exp[-(\tau-u)]\,[1-\Psi_t(u+B)]^K\right]}_{=\,\psi_{t+1}(\tau)\,\text{from
Eq.~\eqref{eq:e8}}},
  \label{eq:e9}
\end{eqnarray}
or in surprisingly simple form,
\begin{eqnarray}
\psi'_{t+1}(\tau)+\psi_{t+1}(\tau)\!\!&=&\!\!- K[1-\Psi_t(\tau+B)]^{K-1}\,\Psi'_t(\tau+B),
  \label{eq:e10}
\end{eqnarray}
which reduces to, using Eq.~\eqref{eq:e5},
\begin{eqnarray}
\Psi''_{t+1}(\tau)+\Psi'_{t+1}(\tau)\!\!&=&\!\!K[1-\Psi_t(\tau+B)]^{K-1}\,\Psi'_t(\tau+B).
  \label{eq:e11}
\end{eqnarray}

\section*{SI B: Asymptotic solution of delay evolution with exponential noise}\label{sec:SIB}

\addcontentsline{toc}{section}{\protect\numberline{}SI B: Asymptotic solution of delay evolution with exponential noise}

\subsection*{Case I: $B>B^*_{\text{c}}$, time-independent stationary solution}

\addcontentsline{toc}{subsection}{\protect\numberline{}Case I: $B>B^*_{\text{c}}$, time-independent stationary solution}

\noindent Let us look for a stationary, i.e. time independent, solution of Eq.~\eqref{eq:e11} at large time.
Such a function $\Psi(\tau)$
satisfies
\begin{eqnarray}
\Psi''(\tau)+\Psi'(\tau)=K[1-\Psi(\tau+B)]^{K-1}\,\Psi'(\tau+B).
  \label{eq:e12}
\end{eqnarray}

Note that in the limit $B\to\infty$ the right-hand side of this equation is equal to $0$. This is trivial, as in that limit the evolution given in Eq.~(2) imposes $\tau_i(t)=\varepsilon_i(t)$, implying that $\Psi(\tau)=e^{-\tau}\mathbf{1}_{\tau>0}$ is the correct time-independent solution.

Another way to see this is to rewrite Eq.~\eqref{eq:e12} as 
\begin{eqnarray}
0=-\frac{\partial}{\partial\tau}\left[
\Psi' +\Psi+\left[1-\Psi\left(\tau + B\right)\right]^K\right],
\end{eqnarray}
which means that:
\begin{eqnarray}
    \Psi' +\Psi+\left[1-\Psi\left(\tau + B\right)\right]^K = c.
\label{eq:FirstOrdPsi}
\end{eqnarray}

Using that $\left[1-\Psi(\tau + B)\right]\approx 1$
for large values of the argument (i.e. large $B$),
this reduces to: 
\begin{eqnarray}
    \Psi' +\Psi = c-1
\end{eqnarray}
defining for convenience $u\equiv\Psi-\Psi(0)$
then we have an inhomogeneous initial value 
problem:
\begin{eqnarray}
    u' +u &&= c-1-\Psi(0)\equiv\gamma \nonumber\\
    u(0) &&=0
\end{eqnarray}
which has the unique solution:
\begin{eqnarray}
u=\Psi-\Psi(0)=\gamma\left[1-e^{-\tau}\right].
\end{eqnarray}

In order for $\Psi\rightarrow 0$ for large values
of its argument, the constant of integration
must be $\gamma=-\Psi(0)$ and therefore
\begin{eqnarray}
    \Psi=\Psi(0)e^{-\tau}.
\end{eqnarray}
We may then restore a finite $B$ in Eq.~\eqref{eq:FirstOrdPsi}. Keeping the lowest-order terms, the correction to the differential equation reads:
\begin{eqnarray}
    \Psi' +\left[1-K e^{-B}\right]\Psi = c-1.
\end{eqnarray}
This correction is important because it shows that this improved, but still approximate, solution behaves as 
$\Psi \propto e^{-\alpha \tau}$ with $\alpha <1$. Therefore, using  $\Psi \propto e^{-\alpha \tau}$ is a reasonable approximation which can be used to obtain further constraints to our problem. Using $\Psi \propto e^{-\alpha \tau}$ in Eq.~\eqref{eq:FirstOrdPsi} and keeping only the lowest-order terms leads to
\begin{eqnarray}
1-\alpha=K\exp(-B\alpha).
  \label{eq:e13}
\end{eqnarray}

\begin{figure}[H]
  \begin{center}
    \includegraphics[width=\textwidth]{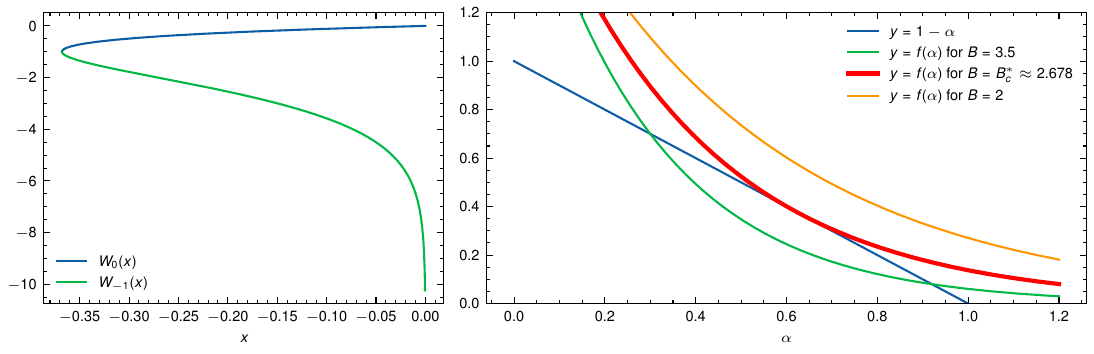}
  \end{center}
  \caption{Left: the two branches of the Lambert function. To recover the right limit $\alpha\to 1$ as $B\to\infty$ for fixed $K$, the $W_0$ branch should be used to find the value of $\alpha_{\text{c}}^*$, while the other branch should be used to obtain the value of $B_{\text{c}}^*$. Right: Plots of $f(\alpha)$ for
    $K=2$ and several values of $B$. The corresponding critical values are $B^*_{\text c}\approx 2.678$ and $\alpha^*_{\text c}\approx 0.627$.\label{lambertfig}}
\end{figure}

A solution of Eq.~\eqref{eq:e13} can only exist if $\alpha<1$ since the right-hand side of the equation, which we denote by $f(\alpha)\equiv Ke^{-B\alpha}$ on Fig.~\ref{lambertfig}, is a positive quantity.

In order to solve Eq.~\eqref{eq:e13} we multiply  each side
by $-B$ and, denoting $w= - B (1-\alpha)$, notice that it takes the form 
\begin{equation} 
w e^w = - B K e^{-B}   \Leftrightarrow  B (1- \alpha ) = W( - B K e^{-B} )   
\label{eq:solualpha} 
\end{equation} 
which determines $\alpha$. Here $W(x)$ is the Lambert function, i.e. the solution of $W e^{W}=x$. No solution exists for $x<-1/e$, while for $x \in [-1/e,0]$ the Lambert function has two branches, shown on the left panel of Figure~\ref{lambertfig}
$W_0(x) \in [0,-1]$ and $W_{-1}(x) \in [-1,-\infty]$. They meet at $W_0(-1/e)=W_{-1}(-1/e)=-1$. Hence, when $B K e^{-B} > 1/e$
there are no solutions for $\alpha$, while for $B K e^{-B} < 1/e$ two solutions appear. 

Varying $B$ at fixed $K>1$, there is thus a transition at $B=B^*_{\text{c}}$, which satisfies
\be 
B^*_{\text{c}} e^{-B^*_{\text{c}}} = \frac{1}{e K} \label{eq:Bc}, 
\ee 
leading in turn to an unique solution for $\alpha$, namely
\begin{equation}\label{eq:e15}
1- \alpha^*_{\text c}=1/B^*_{\text c}.
\end{equation}
Equation~\eqref{eq:Bc} is itself solved using the Lambert function, as $B^*_{\text{c}}= - W(-1/(e K))$. Because $K>1$, two roots are possible, but for $B \gg 1$ the solution should read $\psi(\tau)\approx e^{-\tau}$, meaning that we should {\it decrease} $B$ until we reach the transition. Hence, we must choose
the {\it largest} root of \eqref{eq:Bc}, so that the critical point is given by
\be 
B^*_{\text{c}} = - W_{-1}(-1/(e K)) > 1.
\ee 
The numerical values of $B^*_{\text c}$ and $\alpha^*_{\text c}$ for a few different values of $K$ are shown in Table \ref{table1}.
\begin{table}[H]
\begin{center}
\begin{tabular}{ccc}
\hline\hline\arrayrulecolor{white}\hline\hline$K$&\quad\quad\quad$B^*_{\text c}$\quad\quad\quad\quad&$\alpha^*_{\text c}$\quad\\\arrayrulecolor{black}\hline 2 & 2.67835 & 0.626636\\ 3 & 3.28938 & 0.695991 \\ 4 & 3.69263 & 0.729190 \\ 5 & 3.99431 & 0.749644 \\ 6 & 4.23519 & 0.763883\\ 7 & 4.43557 & 0.774550\\ 8 & 4.60702 & 0.782940 \\ 9 & 4.75680 & 0.789775 \\ 10 & 4.88972 & 0.795489\\\hline
\end{tabular}
\end{center}
\caption{The values of $B^*_{\text c}$ and $\alpha^*_{\text c}$ from Eq. (5) for a few different values of $K$ for the MF case.\label{table1}}
\end{table}

For $B \geq B^*_{\text{c}}$ there is thus a time independent solution
for $\psi(\tau)$. In that phase one has then $B K e^{-B} < 1/e$ and the value of $\alpha$ is given by
\eqref{eq:solualpha} with $W=W_0$ (which has the correct limit $\alpha=1$ for $B \to +\infty$, 
while $W_{-1}$ gives $\alpha=0$). 

\vspace{3mm}
\paragraph{\textbf{Square-root singularity}} we may also see how $\alpha-\alpha^*_{\text c}$ behaves compared with $B-B^*_{\text c}$. Substituting $\alpha = \alpha^*_{\text c}+\eta$ and $B = B^*_{\text c}+\epsilon$ into Eq.\eqref{eq:e13} and keeping terms of first order in $\epsilon$ and up to second order in $\eta$, we obtain

\begin{equation}\label{eq:sqroot_eq1}
\begin{split}
1-\alpha^*_{\text{c}} - \eta &\approx \frac{1}{B^*_{\text{c}}} \exp(-B^*_{\text{c}} \eta - \alpha^*_{\text{c}} \epsilon) \\
&\approx \frac{1}{B^*_{\text{c}}}\left[1-B^*_{\text{c}}\eta - \alpha^*_{\text{c}} \epsilon + \frac{1}{2} \left(B^{*2}_{\text{c}}\eta^2 + \alpha^{*2}_{\text{c}} \epsilon^2\right)\right] \\
&\approx 1-\alpha^*_{\text{c}} - \eta -\frac{\alpha^*_{\text{c}}}{B^*_{\text{c}}} \epsilon  +\frac{1}{2} B^*_{\text{c}} \eta^2.
\end{split}
\end{equation}
which then becomes
\begin{equation}\label{eq:sqroot_eq2}
\eta \approx \sqrt{\frac{2\alpha_{\text c}^*}{2B^{*2}_{\text c}}\epsilon}.
\end{equation}
Finally, this means that $\alpha - \alpha^*_{\text c}\underset{B\to B^*_{\text c}}{\sim} (B-B^*_{\text c})^{1/2}$.

\subsection*{Case II: $B<B^*_{\text{c}}$, travelling front solution}

\addcontentsline{toc}{subsection}{\protect\numberline{}Case II: $B<B^*_{\text{c}}$, travelling front solution}

In addition to the previous case, there can also be a travelling front solution, for which we assume the asymptotic form $\psi(\xi)\equiv\psi_{t\rightarrow\infty}(\xi)\sim\exp(-\alpha\xi)$ for large $\xi\equiv\tau-vt$. This assumption reduces Eq.~\eqref{eq:e10} to 
\begin{eqnarray}
  -\alpha\exp(\alpha v) +\exp(\alpha v)= K\exp(-\alpha B),
  \label{eq:e17}
\end{eqnarray}
and then Eq.~\eqref{eq:e17} can be written as
\begin{eqnarray}
  \frac{1-\alpha}{K}\exp(\alpha v) =  \exp(-B\alpha).
  \label{eq:e18}
\end{eqnarray}
The two types of asymptotic forms, together with the exponential decay assumption of $\psi$ can be consistent with each other if $v=0$ when $B>B^*_{\text c}$.

\vspace{3mm}
\noindent\underline{\bf Case $B<B^*_{\text c}$:}

\vspace{3mm}
\noindent For $B<B^*_{\text c}$, we assume that for large $\xi$,
$\psi(\xi)\sim\exp(-\alpha^*_{\text c}\xi)$, leading to
\begin{eqnarray}
  \frac{1-\alpha^*_{\text c}}{K}\exp(\alpha^*_{\text c}
  v) =  \exp(-B\alpha^*_{\text c}).
  \label{eq:e19}
\end{eqnarray}
Since $\tilde\alpha_{\text c}$ is related to $B^*_{\text c}$ through Eq.~\eqref{eq:e15}, combining with \eqref{eq:e19} leads to $v = B^*_{\text c}-B$.

We close SI B with the note that Eq.~\eqref{eq:FirstOrdPsi} we have analytically solved for $K=2,3,4$ and 5, and note that the expressions for the solutions become increasingly complex with increasing values of $K$.

\subsection*{Finite-$N$ effects\label{sec2c}}

\addcontentsline{toc}{subsection}{\protect\numberline{}Finite-$N$ effects}

\begin{figure}[H]
    \centering    \includegraphics[width=0.5\linewidth]{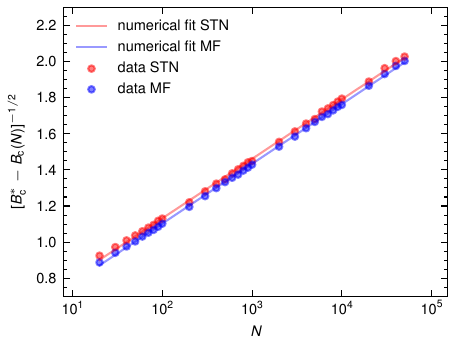}
    \caption{Finite-$N$ effects described by Eq. \eqref{eq:em6} for the critical buffer value $B_{\text c}(N)$ at $K=5$, obtained from the order parameter analysis. The best linear fit values of $(a,b)$ are $\approx(0.4723,0.1432)$ (MF) and $\approx(0.4407,0.1437)$ (STN). The dependence on $\ln N$ implies slow convergence to the asymptotic value $B^*_{\text c}$.}
    \label{fig3}
\end{figure}

From the comparison of Fig. 1(a-b) and Table SI.1 in SI B, it is clear that the values of $B_{\text c}(N)$ and $\alpha_{\text c}(N)$ for $N=10,000$ and $K=5$ do not match the analytical solution $B^*_{\text c}\approx3.99431$ and $\alpha^*_{\text c}\approx0.749644$. We reason in Fig. \ref{fig3} that this is caused by finite-$N$ effects. Indeed, we find that the $B_{\text c}(N)$ data for both the MF and the STN cases can been fitted very well with a function of the type
\begin{eqnarray}
B_{\text c}(N) = B^*_{\text c} -\frac{1}{(a+b\ln N)^2}.
\label{eq:em6}
\end{eqnarray}
One plausible explanation for the $\ln N$-dependence can be the ``small-world phenomenon'': since the inputs of every node comes from $K$ randomly chosen nodes at every time step for the MF case, on average the links among all possible nodes have been established within $\mathcal{O}(\ln N)$ time steps, which is the diameter of a sparse random graph~\cite{Fernholz2007,Chung2001}. The same can be argued for STN as well, but additional constraints (such as clustering within the supply chain, or any sector structure) can make this effect more pronounced. Remarkably, Eq.~\eqref{eq:em6} predicts a $(\ln N)^{-2}$ asymptotic convergence for large $N$, which is strongly reminiscent of the classic result of Brunet \& Derrida \cite{brunet1997shift} for the finite size correction to the velocity of traveling fronts that appear in the analytical solution of our problem on a tree-like structures (SI E and Ref. \cite{derrida1988polymers}). Although the precise mathematical connection is unclear to us at this stage, it is not unreasonable to surmise that such a connection indeed exists. 

\subsection*{Fluctuations in mean delay per node  \label{sec3c}}

\addcontentsline{toc}{subsection}{\protect\numberline{}Fluctuations in mean delay per node}

We also characterized the fluctuations in the mean delay per node close to criticality. For this, we introduced the fluctuations for a run over the time interval $\llbracket 0,T\rrbracket$ as 
\begin{equation}\label{eq:fluctuations}
\Delta\tau(t) \equiv \frac{1}{N}\sum_i \left[\tau_{i}(t) - vt\right]\quad\mbox{for $0\le t\le T$,}
\end{equation}
where $v$ was computed from a linear regression on the mean delay per node $\sum_i\tau_i/N$ data over the time interval $\llbracket0,T\rrbracket$. We then computed the variogram of the fluctuations, defined as
\begin{equation}\label{eq:variogram_def}
\mathcal{V}(\ell) \equiv \text{Var}\left[\Delta\tau(t+\ell)-\Delta\tau(t)\right],
\end{equation}
where the variance is computed over $t$.
\begin{figure}[H]
\centering
\includegraphics[width=0.5\linewidth]{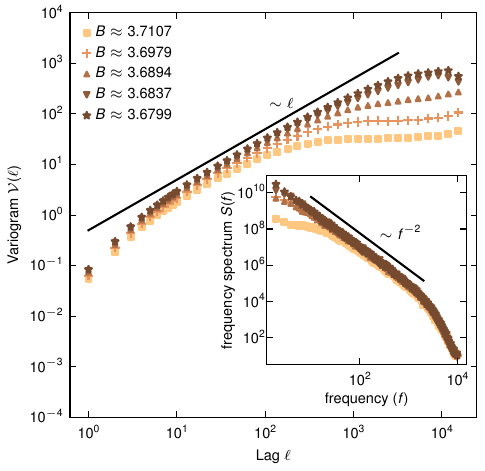}
\caption{Main graph: Variogram of fluctuations in mean delay per node (the same $B$-values used in Fig. 3 in the main text, using also the same color scheme). Inset: corresponding spectral density of fluctuations $S(f)$. See text for details.\label{freq_spec}}
\end{figure}

For purely diffusive processes, $\mathcal{V}(\ell)\propto\ell$. The variogram is plotted in Fig. \ref{freq_spec} (main graph) for several $B$-values above $B_{\text c}(N)$, where we see that the variogram is initially linear in $\ell$, but departs from the linear trend at larger $\ell$-values. Moreover, the point of departure from the linear trend occurs at progressively larger values of $\ell$ as $B$ approaches $B_{\text c}(N)$. In other words, the mean delay per node follows a biased random-walk close to criticality: on top of the average drift $v$, the mean delay per node evolves completely randomly at each time step. This behavior is in line with the spectral density of fluctuations $S(f)$, defined as
\begin{equation}\label{eq:spectral_density}
S(f) \equiv \left\vert \sum_{t=1}^T \Delta \tau(t) e^{-2\iu \pi ft/T} \right\vert^2,
\end{equation} 
becoming closer to $f^{-2}$ for progressively smaller values of $f$ as $B$ approaches $B_{\text c}(N)$ Fig. \ref{freq_spec} (inset). 

\section*{SI C: Real-world data statistics}\label{sec:SIC}

\addcontentsline{toc}{section}{\protect\numberline{}SI C: Real-world data statistics}

The data sets for the high school and  the workplace were taken from \url{sociopatterns.org}~\cite{highschooldata,workplacedata}. In these datasets, agents --- pupils and employees respectively --- were tracked by wearable sensors, and contacts among agents were recorded at 20 second intervals (the time interval between two consecutive time steps is therefore 20 seconds). The time series of the agents'  contacts at discrete time steps yielded the corresponding temporal networks, which provided us with the temporal adjacency matrix ${\mathbf A}(t)$.

As described in the main text associated with Fig. 4, we constructed the ``event representations'' of these temporal networks, with each disjoint network component at every time step defining an event. Because of the ways the sensors work, an (example) event of agents $a_1$, $a_2$ and $a_3$ at some time step may be registered as simultaneous contacts between $a_1$ and $a_2$, and between $a_1$ and $a_3$, while a contact between $a_2$ and $a_3$ is not registered. In order to suit the current purpose however, in the event representation we consider it as an event where all three agents are simultaneously in contact with each other. 

The statistics for the data sets that were used in Fig. shown in Table \ref{tab:RWtable}. [Sparsity in Table \ref{tab:RWtable} is defined as the percentage of agents that participated in single-agent events, summed over all temporal layers. For example, in Fig. \ref{fig:sparse_diagram}(b), the sparsity is obtained as $(0+\frac12+\frac12)/3=1/3\approx33.33\%$.]
\begin{table}[H]
\begin{center}
\begin{tabular}{ccc}
    \hline
        Quantity & \quad\quad Highschool data \quad\quad & \quad\quad Workplace data \quad\quad \\
        \hline
        Total time steps & 8937 & 9679\\
        Unique agents & 328 & 208\\
        Events & 131251 & 33572\\
        \quad Events per timestep\quad & 14.68 $\pm$ 4.30 & 3.47 $\pm$ 2.13\\
        Agents per event & 11.70 $\pm$ 12.59 & 2.45 $\pm$ 0.72\\
        Sparsity & $47.63\%$& $95.90\%$\\
        \hline
    \end{tabular}
    \end{center}
    \caption{Summary statistics of the temporal network data sets from a highschool \cite{highschooldata} and  a workplace \cite{workplacedata}.}\label{tab:datatable}
    \label{tab:RWtable}
\end{table}

\section*{SI D: Effects of heterogeneity and sparsity of STNs on criticality}\label{sec:SID}

\addcontentsline{toc}{section}{\protect\numberline{}SI D: Effects of heterogeneity and sparsity of STNs on criticality}

In order to get an insight into what causes higher-order transitions for real-world temporal networks in contrast to second order transitions for the MF or STN cases [as in Fig. 2(a) in the main text], we simulated delay developments on STNs upon introducing sparsity and heterogeneity into them.

\begin{figure}[!h]
    \centering
    \includegraphics[width=0.99\textwidth]{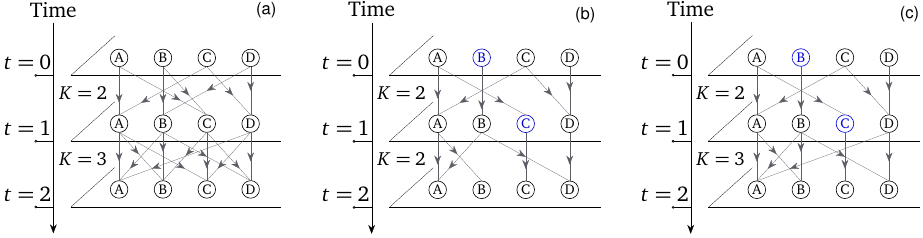}
    \caption{{Schematics for synthetic temporal networks (STNs) with heterogeneity and sparsity. (a) With heterogeneity, but no sparsity: $K$ is the same for all system components per time step, but is different across time steps (for example, $K=1$ and $3$ at $t=0$ and $1$ respectively. (b) With sparsity, but no heterogeneity. Sparse nodes (blue) are created by connecting the system components only to itself between time steps $t$ and $t+1$. (c) A temporal network with both heterogeneity and sparsity, as seen in real-world temporal networks.}
    }\label{fig:sparse_diagram}
\end{figure}

We define heterogeneity in STNs as the value of $K$ being different at different time steps --- albeit the same for all nodes at any given time step [a heterogeneous, but not sparse STN is shown in Fig. \ref{fig:sparse_diagram}(a)]. Sparsity in a STN is defined in line with SI C: it is determined by the average percentage of system components whose delay at any given time step does not influence any other ones in the following time step [a sparse STN without heterogeneity is shown in Fig. \ref{fig:sparse_diagram}(b), the sparsity of this network is $\left(\frac14+\frac14+0\right)/3=1/6\approx16.67\%$ ]. Defining them in this way means that we can separately control heterogeneity and sparsity [and if needed, can also combine them, as shown in Fig. \ref{fig:sparse_diagram}(c)] while synthesizing temporal networks.

\begin{figure}[!h]
\centering
\begin{minipage}{0.49\linewidth}
\centering
\includegraphics[width=\linewidth]{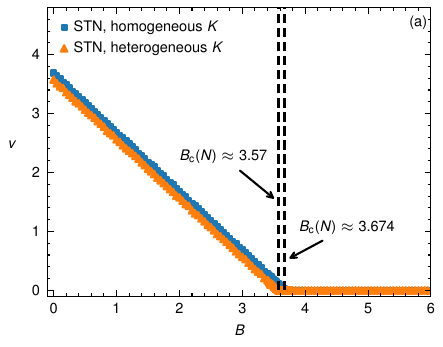}
\end{minipage}
\begin{minipage}{0.49\linewidth}
\centering
\includegraphics[width=\linewidth]{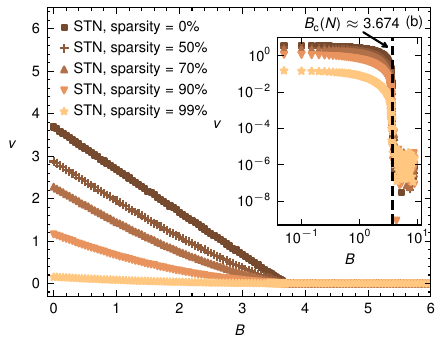}
\end{minipage}
  \caption{Effect of heterogeneity alone (a) and sparsity alone (b) on criticality for $N=10,000$. In panel (a) the order parameter for homogeneous STN with $K=5$ is plotted alongside the one for a heterogeneous STN with \( K \sim \text{Uniform}\{2, 3, \ldots, 8\} \). Heterogeneity alone changes the magnitude of $B_{\text c}(N)$, but does not change the order of transition. In contrast, panel (b) shows that sparsity alone does not change the magnitude of $B_{\text c}(N)$, but pushes the transition to be of higher order. See text for details.}
  \label{fig:sparsity_v_graph}
\end{figure}

The results of our experiments with delay developments on (a) heterogeneous but not sparse and (b) sparse but homogeneous temporal networks for $N=10,000$ are shown in Fig. \ref{fig:sparsity_v_graph}(a) and (b) respectively. In panel (a) the value of $K$ per time step is chosen from a uniform distribution between 1 and 9; we choose this range in order to maintain the average $K$ at 5, allowing us a comparison to Fig. \ref{fig:sparse_diagram}(a) for which $K=5$. Evidently, the transition is still second order, but heterogeneity alone in $K$ lowers the value of $B_\mathrm{c}(N)$ [c.f. \ref{fig:sparsity_v_graph}(a)]. In contrast, sparsity alone does not change the magnitude of $B_{\text c}(N)$ but pushes the transition to be of higher order. 

\section*{SI E: Finite temperatures, continuous tree}\label{sec:SIE}

\addcontentsline{toc}{section}{\protect\numberline{}SI E: Finite temperatures, continuous tree}

The aim of this section is to (a) make a more direct connection with the Directed Polymer problem and provide a generalization of our model to non-zero temperatures that reveals a full line of transition in the plane (buffer amplitude, temperature) -- see Fig. \ref{fig:appendixe_fig}; and (b) give another exact calculation of the properties of the model on continuous tree-like structures, rather than in the mean-field limit considered in the main text. Reassuringly, the results obtained on trees are very similar to those obtained in the mean-field limit.

\subsection*{Finite temperature generalization} 

\addcontentsline{toc}{subsection}{\protect\numberline{}Finite temperature generalization}

To make apparent the connection with the directed polymer (DP) problem, it is interesting to define a ``finite temperature" extension of our model. Introducing the variable $Z_{i}(t) = e^{\beta \tau_{i}(t)}$, where $\beta$ is
the inverse temperature parameter, the recursion is generalized as
\be \label{finitetemp1}
Z_i(t+1) = e^{\beta \epsilon_{i}(t)} \left(1 + e^{-\beta B} \sum_{j \in \partial_i} Z_{j}(t) \right). 
\ee 
where $\partial_i$ denote the set of $K$ sites connected to $i$. When $\beta \to +\infty$ one
recovers the ``zero temperature'' recursion considered in the main text:
\be 
\tau_i(t+1) = \epsilon_{i}(t) + \max\left(0, \max_{j \in \partial_i} (\tau_j(t)-B) \right). 
\ee 
The variable $Z_i(t)$ can be interpreted as a canonical partition function for polymers ending at $i,t$ and the variable $\tau_i(t)= \log Z_i(t)$ as minus the free energy.
Note that at finite temperature the variable $\tau_i(t)$ can become negative. 

\subsection*{Cayley tree}

\addcontentsline{toc}{subsection}{\protect\numberline{}Cayley tree}

We now study the model on the Cayley tree (CT) where each site at generation $t+1$ 
is connected to $K$ sites at generation $t$. Since the $\epsilon_{i}(t)$
are i.i.d.,
assuming that the leaves at $t=0$ to be identical and independent, those at generation $t$ 
are also i.i.d. and one can rewrite \eqref{finitetemp1} as a recursion for a single random variable
$Z(t)$
\be \label{finitetemp}
Z(t+1) = e^{\beta \epsilon(t)} \left(1 + e^{-\beta B} \sum_{j=1}^K  Z^{(j)}(t) \right), 
\ee 
where $Z^{(j)}(t)$ are $K$ independent copies of the random variable $Z(t)$. 
This is a variant of the DPCT problem studied in Ref. \cite{derrida1988polymers}.
Let us introduce, as in that work, the generating function, defined for $\tau \in \mathbb{R}$
\be \label{gener} 
G_t(\tau) = \langle e^{- e^{- \beta \tau} Z(t) } \rangle 
\ee 
with $G_t(\tau) \in [0,1]$, $G_t(-\infty)=0$ and $G_t(+\infty)=1$. From Eq. \eqref{finitetemp}, and since the $Z^{(j)}(t)$ for different $j$ are independent, one immediately obtains the recursion
\be \label{recG} 
G_{t+1}(\tau) = \int_0^{+\infty} \dint\epsilon~ p(\epsilon) e^{- e^{- \beta (\tau-\epsilon)}  }   [ G_t(\tau+B-\epsilon) ]^K,
\ee 
where we denote $p(\epsilon)$ the PDF of $\epsilon(t)$. Since $G_t(\tau)  = \langle e^{- e^{- \beta (\tau- \tau(t))}  } \rangle$
one has
\be 
\lim_{\beta \to +\infty} G_t(\tau) =  \langle \theta(\tau- \tau(t)) \rangle  = {\rm Prob}(\tau(t) < y) = P_t^<(\tau) = \int_0^\tau \dint\tau'~ P_t(\tau') 
= 1 - \Psi_t(\tau).
\ee
Hence in the limit $\beta \to +\infty$ the recursion \eqref{recG} becomes
\be \label{eq:primitive} 
P^<_{t+1}(\tau)  = \int_{0}^\tau \dint\epsilon~ p(\epsilon)  [ P^<_{t}(\tau+B-\epsilon) ]^K.
\ee 
This relation is the primitive of the one in Eq. (3), which can be recovered by taking a derivative of Eq. (47) w.r.t. $\tau$ and using $\frac{\dint}{\dint\tau} P_t^<(\tau) = P_t(\tau) = \psi_t(\tau)$. So we have checked that Eq. (3) is exact on the Cayley tree. 

\subsection*{Continuous tree}

\addcontentsline{toc}{subsection}{\protect\numberline{}Continuous tree}

As in Ref. \cite{derrida1988polymers} it is easier to study the continuum version of the tree. 
So now $t$ is a continuum variable and the tree branches in two with a rate $\varphi dt$. 
We assume that the noise becomes Brownian in the limit, with a drift $\bar \epsilon$ (corresponding to the average delay for individual events in the discrete case)
\be 
\epsilon(t)  \to \bar \epsilon dt + \sigma dW(t), 
\ee 
where $W(t)$ is a standard Brownian motion.
Now $Z(t)$ becomes a continuous process in time and its evolution on a time interval $dt$ is as follows
\bea \label{evolution} 
Z(t+dt) = \left\{\begin{matrix}
Z(t) e^{ \beta (\bar \epsilon dt + \sigma dW(t)) } &\text{with prob.}& \quad 1- \varphi \, dt \\
  1 + e^{-\beta B} \sum_{j=1}^K Z^{(j)}(t)  &\text{with prob.}& \quad \varphi \, dt 
\end{matrix}\right..
\eea 
With the same definition of the generating function $G_t(\tau)$ as in \eqref{gener} 
and using Ito's rule we find that under the evolution \eqref{evolution}
it satisfies the partial differential equation
\be \label{KPPbeta} 
\partial_t G_t(\tau) = - \bar \epsilon \partial_\tau G_t(\tau) + \frac{\sigma^2}{2} \partial^2_\tau G_t(\tau)
+ \varphi ( e^{- e^{- \beta \tau}  }  G_t(\tau+B)^K - G_t(\tau) ), 
\ee 
which is the continuum analog of Eq. \eqref{recG} (with the same boundary conditions).

\vspace{3mm}
\noindent{\bf Moving phases: high and low temperature phases}.
Let us look for a traveling wave solution. We insert
\be 
G_t(\tau) = w(\tau-v t),   \quad \quad w(-\infty)= 0, \quad \quad w(+\infty)=1.
\ee 
One sees that if $v>0$ the term $e^{- e^{- \beta \tau}  } \approx e^{- e^{- \beta v t}  } \approx 1$ in the region of the front
hence one obtains  
\be 
\frac{\sigma^2}{2} w''(y) + (v - \bar \epsilon ) w'(y) + \varphi( w(y+B)^K - w(y) ) = 0.
\ee 
Proceeding as in Ref. \cite{derrida1988polymers} we look for a tail at large $y>0$
\be 
w(y) \approx 1 - e^{- A y}.
\ee 

Collecting the terms proportional to $e^{- A y}$ one obtains the condition which determines
the velocity as a function of $A$
\be \label{vA}
v = {\sf v}(A) := \bar \epsilon + \frac{\varphi(K e^{- A B}-1)}{A} + \frac{\sigma^2}{2} A. 
\ee 
It remains to determine $A$. We will proceed as in \cite{derrida1988polymers}.
First one notes from \eqref{gener} that for large positive $\tau$, 
$G_t(\tau)\approx 1 - e^{-\beta \tau} \langle Z(t) \rangle$.
On the other hand one easily finds from the recursion \eqref{evolution},
or equivalently from \eqref{KPPbeta}, that $\langle Z(t) \rangle \approx e^{\beta v(\beta) t}$
(which is exact for any $\beta$, where the function ${\sf v}(A)$ is defined in \eqref{vA}).
When this ``first moment tail" is also in the front region defines the {\it high temperature phase}.
In this phase $A=\beta$ (i.e. from the boundary condition at $y=+\infty$)
and $v={\sf v}(\beta)$. 

On the other hand the function ${\sf }(A)$ defined in \eqref{vA} has (for $K>1$) a unique minimum 
as a function of $A$, which we denote $A=A_{\text c}$. As $\beta$ increases from zero, the
front velocity $v={\sf v}(\beta)$ decreases in the high temperature phase and eventually
reaches its minimum value. As in \cite{derrida1988polymers} one obtains
that it {\it freezes} at the temperature $\beta_{\text c}=A_{\text c}$ hence 
\be 
v = v_{\text c} = {\sf v}(A_{\text c}) \quad , \quad \beta > \beta_{\text c}=A_{\text c},
\ee 
which defines the low temperature (glass) phase. In that phase $1-G_t(\tau)\approx \tau e^{- \beta_{\text c} \tau}$ 
at large $\tau$ (a much slower decay than the one in the first moment tail, which is now valid
only in the far forward region of the front). The value of $\beta_{\text c}=A_{\text c}$ is determined by
the root of
\be 
(1 + A_{\text c} B) e^{- A_{\text c} B} = \frac{1}{K} + \frac{\sigma^2}{2 K \varphi} A_{\text c}^2.
\ee 
The r.h.s is an increasing function of $A_{\text c}$ starting from $1/K$ for $A_{\text c}=0$, whereas the
l.h.s. is a decreasing function of $A_{\text c}$ (one has $(1+x)e^{-x} = 1- \frac{x^2}{2} + O(x^3)$ at small $x$),
hence there is a unique root. The corresponding value of the velocity in the low temperature phase is 
\be
v_{\text c}= \bar \epsilon + \frac{\sigma^2}{2} A_{\text c}  + \frac{\varphi(K e^{- A_{\text c} B}-1)}{A_{\text c}}. 
\ee 

Note that for $B=0$ (in the absence of buffers) one has 
\be 
{\sf v}(A) := \bar \epsilon + \frac{\varphi(K-1)}{A} + \frac{\sigma^2}{2} A \quad , \quad 
\beta_{\text c}=A_{\text c}=\sqrt{ \frac{2 \varphi}{\sigma^2} (K-1)} \quad , \quad v_{\text c} = \bar \epsilon  + 2 \sqrt{ \frac{\varphi \sigma^2}{2} (K-1)} .
\ee 

\vspace{3mm}
\noindent{\bf Pinned phase}.
The above analysis is correct only whenever the front velocity $v>0$. When the velocities determined above, i.e. $v={\sf v}(\beta)$ for $\beta < \beta_{\text c}$ and $v=v_{\text c}$ for $\beta>\beta_{\text c}$, vanish then one enters the ``pinned'' phase where $G_t(\tau)$ reaches a stationary limit $G_\infty(\tau)$ for $t \to +\infty$. This corresponds to the situation where delays do not accumulate, and requires that $B$ increases beyond a threshold value $B_{\text c}^*$, which however has a different expression depending on whether $\beta<\beta_{\text c}$ and $\beta>\beta_{\text c}$. Hence there are now two phase boundaries to the pinned phase.
\begin{figure}[H]
  \centering
\includegraphics[width=0.6\textwidth]{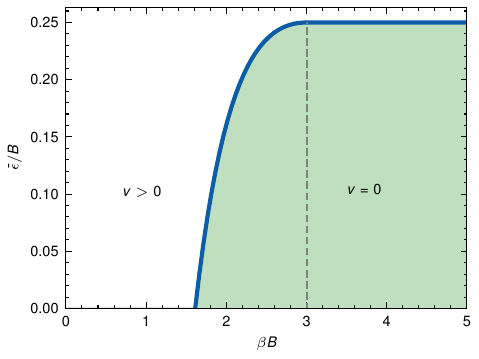}
  \caption{Phase diagram of the model in the plane (inverse temperature $\beta B$, inverse buffer strength ${\bar \epsilon}/{B}$), in the limit where $B$ and $\bar \epsilon$ are both large, and $\varphi=1$. Note that for low enough temperature, { i.e. for $\beta > \beta_{\text c} = r_{\text c}/B$}, 
  the critical buffer strength $B_{\text c}^*$ is independent of temperature. For $\beta < \beta_{\text c} = r_{\text c}/B$, on the other hand, the critical buffer size is temperature dependent. 
  {For $\beta B <  \log K$ only the ``depinned'' phase $v > 0$ exists for $\bar \epsilon>0$}. { Here we have chosen $K=5$ for which $r_{\text c} \approx 3$ and $\log K \approx 1.61$}.}
  \label{fig:appendixe_fig}
\end{figure}

The expression of the phase transition boundary considerably simplifies in the limit $\bar \epsilon \to \infty$, $B \to \infty$ and $\beta \to 0$ with $\bar \epsilon/B$ and $\beta B$ remaining $O(1)$. { In that limit the terms containing the noise amplitude $\sigma$ become subdominant}. We then get that 
$A_{\text c} B = r_{\text c}$ with $r_{\text c}$ given by the solution of 
\[ 
(1 + r_{\text c}) e^{-r_{\text c}} = \frac1K.
\]
For $K=5$, for example, the numerical value of $r_{\text c}$ is very close to $3$. The phase boundary can the be computed to be:
\begin{equation}
    \begin{cases}
        \frac{\bar \epsilon}{B} &= \frac{\varphi}{1 + r_{\text c}}  \qquad (r \geq r_{\text c}) \\ \nonumber 
        \frac{\bar \epsilon}{B} &= { \frac{\varphi}{r} (1 - K e^{-r})^+}=\frac{\varphi}{r} \left[1 - K^{1-r/r_{\text c}} (1+r_{\text c})^{-r/r_{\text c}}\right]^+ \qquad (r \leq r_{\text c}) 
    \end{cases},
\end{equation}
where $r := \beta B$. { Here we restrict to $\bar \epsilon>0$, in the spirit of the model in the main text}.
The corresponding phase diagram in the plane $r, {\bar \epsilon}/{B}$ is given in Fig. \ref{fig:appendixe_fig}. One sees that the zero temperature phase transition found in the main text survives at non zero temperatures, but disappears at high temperatures where only the ``depinned'' phase $v > 0$ exists. 

The phase diagram looks qualitatively similar for values of $B$ and $\bar \epsilon$ that are not infinitely large. Also, one can check that the behaviour of $v$ when $B$ is close to $B_{\text c}^*$ (when it exists) is always linear, i.e. $v \propto B_{\text c}^* - B$. 

\section*{Data and codes for figures}

\addcontentsline{toc}{section}{\protect\numberline{}Data and codes for figures}

\subsection*{For figures in the main text}

\addcontentsline{toc}{subsection}{\protect\numberline{}For figures in the main text}

Below we list the data and the codes to plot the figures.

\begin{itemize}
\item Fig. 2(a): 
\href{https://github.com/jose-moran/timeliness_criticality/tree/main/figures/fig2_critical_buffer/data}{data}, \href{https://github.com/jose-moran/timeliness_criticality/blob/main/figures/fig2_critical_buffer/fig2.ipynb}{code}
\item Fig. 2(b): \href{https://github.com/jose-moran/timeliness_criticality/tree/main/figures/fig2_critical_buffer/data}{data} \href{https://github.com/jose-moran/timeliness_criticality/blob/main/figures/fig2_critical_buffer/fig2.ipynb}{code}
\item Fig. 2(c): \href{https://github.com/jose-moran/timeliness_criticality/tree/main/figures/fig2_critical_buffer/data}{data}, \href{https://github.com/jose-moran/timeliness_criticality/blob/main/figures/fig2_critical_buffer/fig2.ipynb}{code}
\item Fig. 2(d): \href{https://github.com/jose-moran/timeliness_criticality/tree/main/figures/fig2_critical_buffer/data}{data}, \href{https://github.com/jose-moran/timeliness_criticality/blob/main/figures/fig2_critical_buffer/fig2.ipynb}{code}
\item Fig. 3(a): \href{https://github.com/jose-moran/timeliness_criticality/tree/main/figures/fig3a_correlation_collapse/data}{data}, \href{https://github.com/jose-moran/timeliness_criticality/blob/main/figures/fig3a_correlation_collapse/correlation_collapse_fig.ipynb}{code}
\item Fig. 3(b): \href{https://github.com/jose-moran/timeliness_criticality/tree/main/figures/fig3b_persistence_avalanche/data}{data}, \href{https://github.com/jose-moran/timeliness_criticality/tree/main/figures/fig3b_persistence_avalanche/persistence_avalanche_fig.ipynb}{code}
\item Fig. 3(c): \href{https://github.com/jose-moran/timeliness_criticality/tree/main/figures/fig3c_avalanche_mass/data}{data}, \href{https://github.com/jose-moran/timeliness_criticality/tree/main/figures/fig3c_avalanche_mass/avalanche_mass_fig.ipynb}{code}
\item Fig. 4(b): \href{https://github.com/jose-moran/timeliness_criticality/tree/main/figures/fig4b_RW_networks/data}{data}, \href{https://github.com/jose-moran/timeliness_criticality/tree/main/figures/fig4b_RW_networksRW_v_fig.ipynb}{code} 
\item Figure in Methods: \href{https://github.com/jose-moran/timeliness_criticality/tree/main/figures/fig5_single_avalanche/data}{data}, \href{https://github.com/jose-moran/timeliness_criticality/tree/main/figures/fig5_single_avalanche/build_single_avalanche_fig.ipynb}{code}
\end{itemize} 

\subsection*{For figures in the Supplementary Information}

\addcontentsline{toc}{subsection}{\protect\numberline{}For figures in the Supplementary Information}

\begin{itemize}
\item Fig. \ref{fig3}: \href{https://github.com/jose-moran/timeliness_criticality/tree/main/figures/SI.2_finite_size/data}{data}, \href{https://github.com/jose-moran/timeliness_criticality/tree/main/figures/SI.2_finite_size/Finite_size_fig.py}{code}
\item Fig. \ref{freq_spec} (including inset): \href{https://github.com/jose-moran/timeliness_criticality/tree/main/figures/SI.3_f_noise/data}{data}, \href{https://github.com/jose-moran/timeliness_criticality/tree/main/figures/SI.3_f_noise/variogram_fig.ipynb}{code}
\item Fig. \ref{fig:sparsity_v_graph}(a) \href{https://github.com/jose-moran/timeliness_criticality/tree/main/figures/SI.5a_heterogeneous_RW/data}{data}, \href{https://github.com/jose-moran/timeliness_criticality/tree/main/figures/SI.5a_heterogeneous_RW/heterogeneity_fig.ipynb}{code}
\item Fig. \ref{fig:sparsity_v_graph}(b), \href{https://github.com/jose-moran/timeliness_criticality/tree/main/figures/SI.5a_sparse_RW/data}{data}, \href{https://github.com/jose-moran/timeliness_criticality/tree/main/figures/SI.5a_sparse_RW/sparsity_fig.ipynb}{code}
\end{itemize}

\section*{References}

\addcontentsline{toc}{section}{\protect\numberline{}References}